# Visualizing electrostatic gating effects in two-dimensional heterostructures

Paul V Nguyen,[1] Natalie C Teutsch,[2] Nathan Wilson,[1] Joshua Kahn,[1] Xue Xia,[2] Viktor Kandyba,[3] Alexei Barinov,[3] Gabriel C Constantinescu,[4] Nicholas D M Hine,[2] Xiaodong Xu,[1,5*] David H Cobden,[1*] Neil R Wilson[2*]

[1]Department of Physics, University of Washington, Seattle, Washington 98195, USA
[2]Department of Physics, University of Warwick, Coventry, CV4 7AL, UK
[3]Elettra - Sincrotrone Trieste, S.C.p.A., Basovizza (TS), 34149, Italy
[4]TCM Group, Cavendish Laboratory, University of Cambridge, 19 JJ Thomson Avenue, Cambridge CB3 0HE, UK
[5]Department of Material Science and Engineering, University of Washington, Seattle, Washington 98195, USA

**Abstract**
**The ability to directly observe electronic band structure in modern nanoscale field-effect devices could transform understanding of their physics and function. One could, for example, visualize local changes in the electrical and chemical potentials as a gate voltage is applied. One could also study intriguing physical phenomena such as electrically induced topological transitions and many-body spectral reconstructions. Here we show that submicron angle-resolved photoemission (μ-ARPES) applied to two-dimensional (2D) van der Waals heterostructures affords this ability. In graphene devices, we observe a shift of the chemical potential by 0.6 eV across the Dirac point as a gate voltage is applied. In several 2D semiconductors we see the conduction band edge appear as electrons accumulate, establishing its energy and momentum, and observe significant band-gap renormalization at low densities. We also show that μ-ARPES and optical spectroscopy can be applied to a single device, allowing rigorous study of the relationship between gate-controlled electronic and excitonic properties.**

Angle resolved photoemission spectroscopy (ARPES), in which the energy and momentum of photoemitted electrons are measured from a sample subjected to a spectrally narrow ultraviolet or X-ray excitation, is a powerful technique that yields the momentum-dependent single-electron band structure and chemical potential in a solid with essentially no assumptions. It probes only electron states near the surface, and so cannot be applied to conventional semiconductor devices. It is, however, very effective when applied to 2D materials and has been used extensively to study the bands in graphene[1], monolayer transition metal dichalcogenides[2–7], and others[8,9]. Furthermore, μ-ARPES (with a micron-scale beam spot) can be performed[10] on 2D heterostructures (2DHSs)[11] made of stacked exfoliated 2D materials[12–14], suggesting the possibility of monitoring electronic structure during actual device operation. We demonstrate here that momentum-resolved electronic spectra can indeed be obtained during reversible electrostatic gating, enabling direct visualization of chemical potential shifts and band structure changes controlled by the gate electric field.

A limitation of ARPES is that it probes only occupied electron states, and so a semiconductor must first be electron-doped in order to obtain a signal from the conduction band. The usual approach is to deposit alkali metal atoms[1–7,15] which act as an n-type dopant, but this has several limitations: the density cannot be controlled accurately; it can only be reversed by high-temperature annealing; it introduces disorder through the random positions of the dopants; and it chemically perturbs the electronic structure in ways that are hard to calculate. Electrostatic doping has none of these disadvantages, and the accessible carrier densities are most relevant to practical devices.

We first validate our technique using graphene, and then go on to apply it to the 2D transition metal dichalcogenide (TMD) semiconductors, which are of great interest for valleytronics and other applications[16,17]. Although it is widely thought that all the monolayer TMD semiconductors have a direct band gap at the corner $K$ of the hexagonal Brillouin zone, the magnitude of the band gap is hard to determine, as illustrated by the wide range of reported values for monolayer WSe$_2$ (from 1.4 to 2.2 eV [15,18–22]). Also unclear is when the additional conduction band minimum at the lower-symmetry momentum point $\boldsymbol{Q}$ comes into play[19,23]. Using just electrostatic doping, in all the monolayer semiconductors MoS$_2$, MoSe$_2$, WS$_2$ and WSe$_2$, we confirm that the CBE is at $\boldsymbol{K}$, and in each case we obtain a measure of the band gap. We also study the layer-number dependence: in monolayer WSe$_2$ we find that the conduction band minimum at $\boldsymbol{Q}$ is about 30 meV higher than that at $\boldsymbol{K}$, but in bilayer and trilayer WSe$_2$ it is lower than at $\boldsymbol{K}$. In addition, we investigate renormalization of the band structure on gating, and discover that the band gap decreases substantially at moderate electrostatic doping levels.

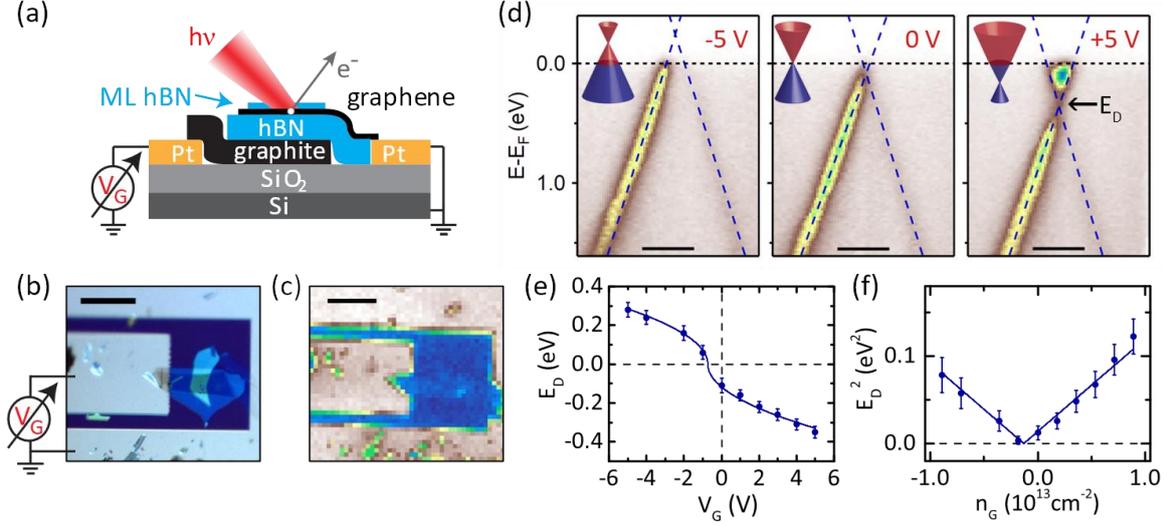

**Figure 1. Visualizing electrostatic doping of monolayer graphene.** (**a**) Schematic of a 2D heterostructure comprising hBN-encapsulated graphene with a graphite back gate. Photoemission is measured with a micron-size beam spot. (**b**) Optical and (**c**) SPEM images of a device ($d_{BN} = 14 \pm 1$ nm). Scale bars are 50 μm. (**d**) Energy-momentum slices near the graphene K-point at gate voltages $V_G$ as labelled. Scale bars are 0.3 Å$^{-1}$. The dashed lines are linear fits; the Dirac point energy $E_D$ is deduced from their crossing point. (**e**) Gate dependence of $E_D$, showing a full swing of about 0.6 eV. (**f**) Plotting $E_D^2$ vs gate-induced doping $n_G$ gives straight lines, consistent with linear dispersion (see text).

## 1. Electrostatic doping of graphene

We first present measurements demonstrating gate-doping of monolayer graphene. The graphene is capped by monolayer hBN and supported on hBN dielectric over a graphite gate, as shown schematically in Fig. 1a. This 2DHS is located in a gap between two platinum electrodes on a SiO$_2$/Si substrate chip. The gate contacts the smaller electrode, as indicated in the optical image in Fig. 1b (see Methods for details). The graphene contacts the larger electrode, which covers most of the silicon chip and is grounded to minimize electrostatic distortion of the photoelectron spectrum when applying a gate voltage. A similar structure with additional contacts to the graphene would function as a high-mobility graphene transistor[24]. After mounting in the chamber, at a stage temperature of 105 K, the 2DHS is located by scanning photoemission microscopy (SPEM; see Methods): an SPEM image of the same area as seen in the optical image is shown in Fig. 1c. For momentum-resolved spectroscopy the ~1 μm excitation beam spot is fixed at a point on the 2DHS.

Fig. 1d shows $(E, k)$ slices near the graphene zone corner $\boldsymbol{K}$, acquired at gate voltages $V_G$ of ±5 V. Following standard practice, we determine the photoelectron energy $E$ which corresponds to electrons removed at the chemical potential $E_F$ by fitting the Fermi-Dirac distribution to the drop in photoemitted intensity around $E_F$, and plot the resulting binding energy $E - E_F$. As expected, the bands show almost linear dispersion, consistent with a section though a Dirac cone centered at $\boldsymbol{K}$. Also as expected, changing $V_G$ shifts the Dirac point energy $E_D$ relative to $E_F$. At $V_G = -5$ V the graphene is p-doped and $E_D > E_F$; at +5 V it is n-doped and $E_D < E_F$. Note that the intensity near $E_D$ is weak because these $E-k$ slices do not pass exactly through $\boldsymbol{K}$, and that the much lower intensity on one side of the cone than the other results from destructive interference between the two carbon sublattices[25] (the slices are close to the $\boldsymbol{\Gamma}-\boldsymbol{K}$ direction).

Fitting a linear dispersion of the form $E(\boldsymbol{k}) = E_D \pm \hbar v_F k$ (dashed lines in Fig. 1d) gives $E_D$ and the Fermi velocity $v_F$. The obtained value of $v_F$ is $(9.3 \pm 0.1) \times 10^5$ ms$^{-1}$ at $V_G = 0\ V$ with very weak $V_G$ dependence (see SI section S2), consistent with the literature and giving us confidence that perturbation of the photoelectron trajectories by the gate electric field is negligible. The gate dependence of $E_D$ is shown in Fig. 1e. For $k_B T \ll E_D$ (reasonable here since $k_B T = 9$ meV) one expects $\pi^{-1} E_D^2/(\hbar v_F)^2 = |n_0 + n_G|$, where[26] $n_0$ is areal electron density in the graphene at $V_G = 0$ and $n_G$ is the gate-induced change. Corrections due to the electron compressibility and resistive potential drops are minor (see SI section S3) and hence $n_G \approx CV_G$ where $C = \epsilon_{BN}\epsilon_0/d_{BN}$ is the geometric areal capacitance of the hBN dielectric with thickness $d_{BN} = 14 \pm 1$ nm and dielectric constant $\epsilon_{BN}$. With $v_F$ and $E_D$ determined from the dispersion, the only unknowns are $\epsilon_{BN}$ and $n_0$. Plotting $E_D^2$ vs $n_G$, as in Fig. 1f, shows the expected linear dependence, with the best fit (solid line) giving $\epsilon_{BN} = 4.5 \pm 0.1$, consistent with the literature[27–29], and $n_0 = (1.8 \pm 0.1) \times 10^{12}$ cm$^{-2}$. We conclude that the technique yields accurate local electronic spectra in the presence of electrostatic gating, as desired.

## 2. Electrostatic doping of 2D semiconductors

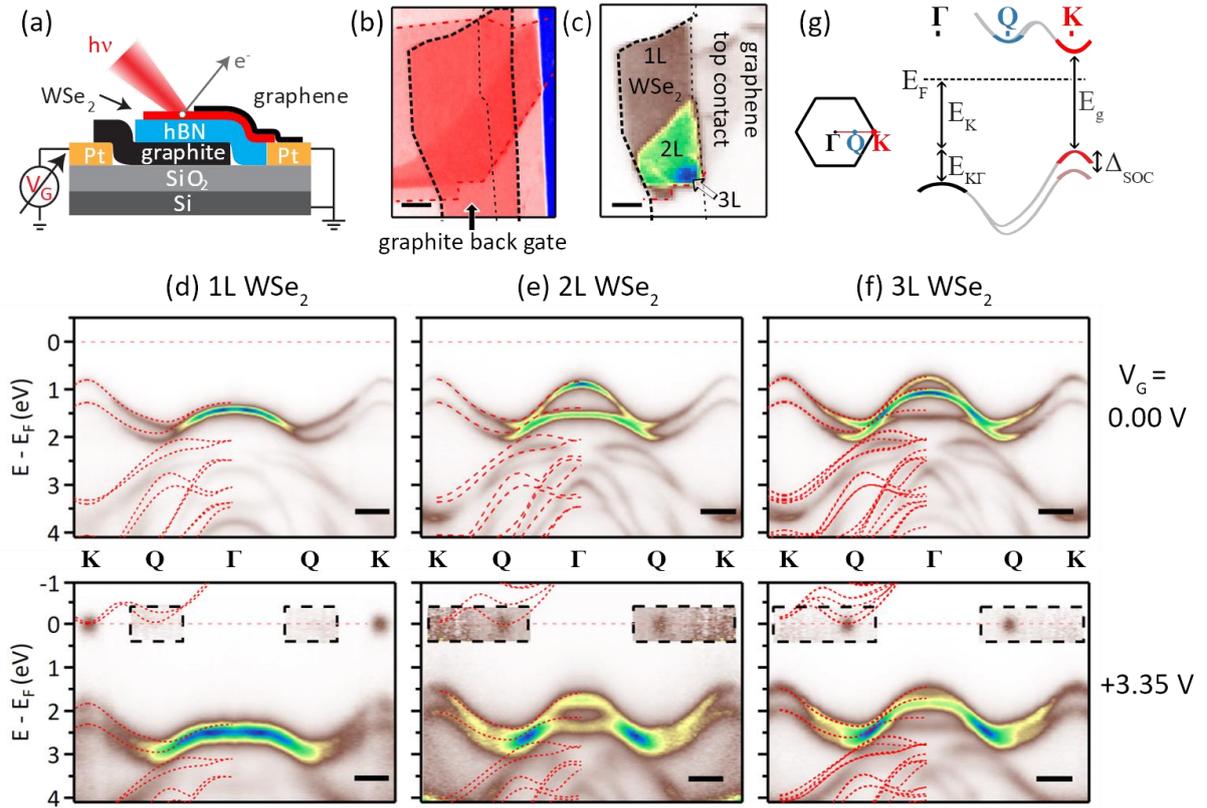

**Figure 2. Layer-number dependent conduction band edge in WSe$_2$.** (**a**) Schematic device, with graphene contact grounded and a gate voltage applied to the graphite back gate. (**b**) Optical and (**c**) SPEM images of a WSe$_2$ device (Device 1, $d_{Bn} = 7.4 \pm 0.5$ nm) with 1L, 2L and 3L regions identified. Scale bars are 5 μm. (**d**)-(**f**) Energy-momentum slices along $\boldsymbol{\Gamma} - \boldsymbol{K}$ for 1L, 2L, and 3L regions respectively. In each case the upper panel is at $V_G = 0$ and the lower at $V_G = +3.35\ V$. Scale bars are 0.3 Å$^{-1}$. The data have been reflected about $\boldsymbol{\Gamma}$ to aid comparison with DFT predictions (red dashed lines). The intensity in the dashed boxes is multiplied by 20 to enhance the weak $\boldsymbol{Q}$-point features. (**g**) Brillouin zone of MX$_2$, and schematic of bands along $\boldsymbol{\Gamma} - \boldsymbol{K}$ showing definitions of energy parameters discussed in the text.

To obtain photoemission spectra from a gated TMD semiconductor, we incorporate an exfoliated MX$_2$ flake in the 2DHS on top of the hBN, partially overlapped by a graphene piece to act as a contact (Fig. 2a). Figures 2b and c are optical and SPEM images of a device with a WSe$_2$ flake that includes monolayer (1L), bilayer (2L) and trilayer (3L) regions. Figures 2d-f show $\Gamma - K$ momentum slices obtained with the beam spot respectively on the 1L, 2L and 3L regions, where $K$ is now the corner of the WSe$_2$ Brillouin zone, shown in Fig. 2g. The measurements in the upper row were made at $V_G = 0$. Here, as expected, only the valence bands are seen. Their evolution with layer number is consistent with the literature[30] and matches the overlaid density functional theory (DFT) predictions quite well.

The measurements in the lower row were made at $V_G = +3.35$ V. In each there is now a signal near $E_F$, implying that the CBE is populated. The contrast has been increased in the dashed rectangles to enhance this signal, which is a resolution-limited spot. Note that all features are broader than at $V_G = 0$, probably because of inhomogeneity in the electrostatic potential. In 1L the signal is at $K$, but in 2L and 3L it is at $Q$, roughly midway between $\Gamma$ and $K$ (see Fig. 2g). This is consistent with evidence from photoluminescence[23] that the gap in the monolayer is direct at $K$ while for 2+ layers it is indirect. The spectra in Figs. 2d-f reveal that in 2 layers the indirect gap is between the CBE at $Q$ and the valence band edge (VBE) at $K$, while for 3+ layers the VBE is at $\Gamma$.

Table 1 displays the band structure parameters for 1–3 L WSe$_2$ and for the other monolayer MX$_2$ species derived[10] from these measurements and from similar ones made on other devices (see SI section S4). In each case for $V_G = 0$ we give the hole effective mass $m_K^*$, valence band edge $E_K$, spin-orbit splitting $\Delta_{SOC}$ at $K$, and difference $E_{K\Gamma}$ between the valence band edges at $K$ and $\Gamma$. We also give the band gap, defined by $E_g = E_C - E_K$, obtained at a gate voltage such that $n_G$ is near $10^{13}$ cm$^{-2}$ (also shown). The energy $E_C$ of the CBE is estimated by assuming a degenerate Fermi gas (see Methods), which for $n_G = 10^{13}$ cm$^{-2}$ yields $E_F - E_C \sim 30$ meV (smaller than the energy broadening in the spectra). With increasing number of layers we discern a small but systematic increase in $\Delta_{SOC}$ and decrease in $m_K^*$ in addition to the substantial decrease in $E_g$ that accompanies the transition from direct to indirect gap. In all the monolayer species the gap is direct at $K$, and the parameters follow the expected trends with composition[31], namely, $E_g$ depends more on the chalcogenide while $\Delta_{SOC}$ depends more on the transition metal.

| | $\Delta_{SOC}$ (eV) | $E_K$ ($V_G = 0$) (eV) | $E_{K\Gamma}$ ($V_G = 0$) (eV) | $m_K^*/m_e$ | $E_g$ (eV) | |
|---|---|---|---|---|---|---|
| 1L MoS$_2$ | 0.17 ± 0.04 | 1.93 ± 0.02 | 0.14 ± 0.04 | 0.7 ± 0.1 | 2.07 ± 0.05 | ($n_G = 12.0 \pm 0.8 \times 10^{12}$ cm$^{-2}$, T = 105 K) |
| 1L MoSe$_2$ | 0.22 ± 0.03 | 1.04 ± 0.02 | 0.48 ± 0.03 | 0.5 ± 0.1 | 1.64 ± 0.05 | ($n_G = 10.4 \pm 0.7 \times 10^{12}$ cm$^{-2}$, T = 105 K) |
| 1L WS$_2$ | 0.45 ± 0.03 | 1.43 ± 0.02 | 0.39 ± 0.02 | 0.5 ± 0.1 | 2.03 ± 0.05 | ($n_G = 9.4 \pm 0.6 \times 10^{12}$ cm$^{-2}$, T = 105 K) |
| 1L WSe$_2$ | 0.485 ± 0.010 | 0.80 ± 0.01 | 0.62 ± 0.01 | 0.42 ± 0.05 | 1.79 ± 0.03 | ($n_G = 8.7 \pm 0.4 \times 10^{12}$ cm$^{-2}$, T = 100 K) |
| 2L WSe$_2$ | 0.501 ± 0.010 | 0.75 ± 0.01 | 0.14 ± 0.01 | 0.41 ± 0.05 | 1.51 ± 0.03 * | ($n_G = 7.1 \pm 0.5 \times 10^{12}$ cm$^{-2}$, T = 100 K) |
| 3L WSe$_2$ | 0.504 ± 0.010 | 0.74 ± 0.01 | 0.00 ± 0.01 | 0.40 ± 0.05 | 1.46 ± 0.03 * | ($n_G = 8.7 \pm 0.4 \times 10^{12}$ cm$^{-2}$, T = 100 K) |

*indirect with CBM at $Q$ and VBM at $K$

**Table 1. Band structure parameters of MX$_2$ semiconductors.** As defined in Fig. 2g, $\Delta_{SOC}$ is the spin-orbit splitting of the valence band at $K$; $E_K$ is the valence band edge at $V_G = 0$; $E_{K\Gamma} = E_K - E_\Gamma$ is the difference between the valence band edges at $K$ and $\Gamma$ at $V_G = 0$; and $E_g$ is the band gap. $m_K^*$ is the effective mass of the valence band edge at $K$ (measured in the $\Gamma$ to $K$ direction using data within 0.1 Å$^{-1}$ of $K$) in units of free electron mass $m_e$. The gate-induced carrier density $n_G$, slightly different in each case, is also given. All samples are on hBN substrates in vacuum.

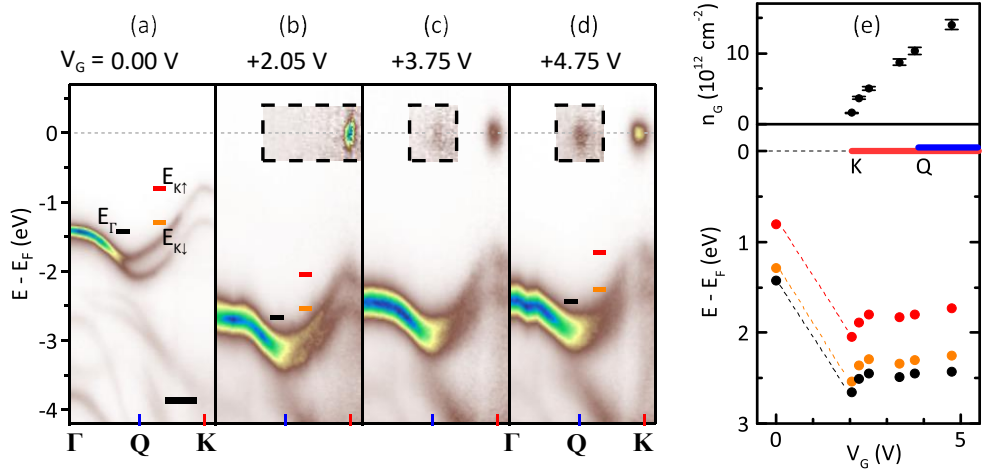

**Figure 3. Band renormalization with carrier density in monolayer WSe$_2$.** (**a**)-(**d**) Energy-momentum slices along **Γ**-**K** for 1L WSe$_2$ at a series of gate voltages in the monolayer part of Device 1. Scale bar is 0.3 Å$^{-1}$. The intensity in the dashed boxes has been multiplied by 20 in b and 40 in c and d. (**e**) Dependence of properties on $V_G$. Bottom panel: upper ($E_{K\uparrow}$, red) and lower ($E_{K\downarrow}$, orange) valence bands at **K** and the valence band maximum at **Γ** ($E_\Gamma$, black). The ranges where the CBE at **K** and **Q** can be seen are indicated by red and blue lines respectively. Top panel: electron doping, calculated using the geometric capacitance and the measured potential drop.

## 3. Doping dependence of electronic structure

Figures 3a-d show spectra from 1L WSe$_2$ at selected gate voltages. In Fig. 3e the bottom panel shows the variation of the valence band features (indicated by color-coded marks) with $V_G$, derived from a series of such measurements. As $V_G$ increases, below the threshold for electron accumulation the bands initially shift downwards, tracking the unscreened gate potential. The top panel shows the corresponding doping level, $n_G$, calculated using the geometric capacitance and accounting for the *measured* potential drop (see SI section S3). Interestingly, above threshold ($V_g \approx 1.5$ V) the valence bands begin to shift *upwards*, implying that the band gap decreases. The conduction band is visible at **K** for $n_G > \sim 10^{12}$ cm$^{-2}$ (indicated by a red line at $E_{bin} = 0$) and becomes visible also at **Q** for $n_G > \sim 10^{13}$ cm$^{-2}$ (blue line), when $E_K$ is about 30 meV below $E_F$. The shape of the valence bands remains the same as they shift upwards at doping levels $n_G < 10^{13}$ cm$^{-2}$, with $\Delta_{SOC}$ and $E_{K\Gamma}$ essentially constant, so it is likely the conduction bands also remain rigid under these conditions. We conclude that that the band minimum at **Q** is higher than that at **K** in monolayer WSe$_2$, by roughly 30 meV when $n_G > \sim 10^{13}$ cm$^{-2}$. We note that scanning tunnelling spectroscopy[19] also indicates that for 1L WSe$_2$ the conduction band minimum at **Q** is very close to that at **K**.

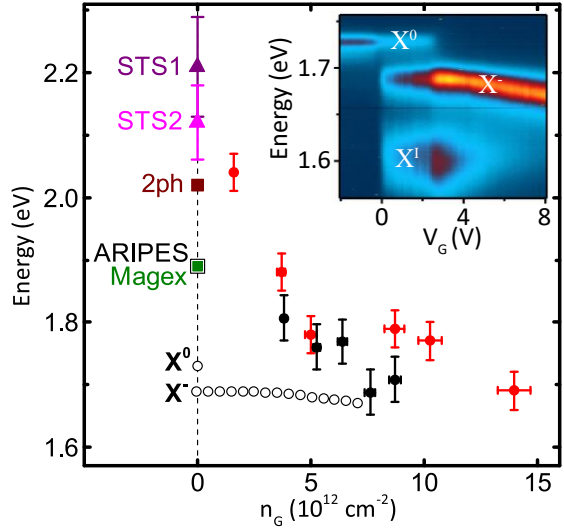

**Figure 4. Carrier-density dependent band gap measurements on monolayer WSe₂.** Solid circles: ARPES measurements of the band gap $E_g$ at 100 K in Device 1 (red) and Device 2 (black, $d_{BN} = 24.5 \pm 0.5$ nm). Open circles: photoluminescence peak positions for the neutral exciton ($X^0$) and negative trion ($X^-$) in Device 2, also at 100 K. The inset shows the raw photoluminescence data; a broad impurity-bound exciton feature $X^I$ is seen too. Also plotted are band gap measurements, nominally at $n_G = 0$, obtained by other techniques, as follows: STS1[18] (purple triangle) and STS2[19] (pink triangle) are from scanning tunneling spectroscopy measurements, monolayer WSe₂ on graphite at T= 4.5 K and 77 K respectively; 2ph (brown square) is from two-photon absorption[20], on SiO₂ at 300 K; ARIPES (black open square) is from inverse photoemission[21], on sapphire at 300 K; and Magex (green solid square) is from magneto-optical measurements[22], encapsulated in hBN at 4 K.

In comparing features between photoemission, tunneling, optical and transport measurements, it is a challenge to account for differences in sample quality, dielectric environment, gate voltage and temperature[32–36]. An advantage of our technique is that these differences can be eliminated, since the same devices are also suitable for optical spectroscopy, and in principle for transport studies, which can be done under exactly the same conditions as the ARPES measurements. Figure 4 shows both the ARPES determination of $E_g$ (black filled circles) and the photoluminescence peak positions (black empty circles), $E_{X^0}$ and $E_{X^-}$, for neutral (X⁰) and charged (X⁻) excitons obtained from a single device (Device 2) as a function of gate doping at 100 K. The values of $E_g$ from Device 1 (red solid circles) are in agreement, to within uncertainty. It is clear that $E_g$ decreases systematically, by ~400 meV, as $n_G$ rises to around $10^{13}$ cm⁻². This behavior resembles the renormalization of the band gap due to free-carrier screening predicted in Ref. [35].

Also plotted are values of the band gap at $n_G = 0$ inferred from scanning tunneling spectroscopy (STS)[18,19], two-photon absorption[20], inverse photemission[21], and magnetooptic[22] studies. An extrapolation of the ARPES $E_g$ measurements to $n_G = 0$ is most consistent with the STS measurements, and comparison with $E_{X^0}$ provides strong support for the widely accepted claim that the binding energy of neutral excitons is several hundred meV. On doping, $E_g$ decreases much faster than $E_{X^-}$, implying dramatic weakening of the exciton binding due to free-carrier screening[34]. The even smaller $E_g$ values reported earlier in monolayers highly doped with alkali metals (down to 1.4 eV for 1L WSe₂) are consistent with a continuation of this renormalization process to higher $n_G$[2,15].

This capability to measure changes in the electron band structure in operating devices opens up many exciting possibilities. For example, in multi-layer 2DHS devices it could be used to study the perpendicular electric-field dependence of the bands[3,37], including potentially inducing band inversion and consequent topological phase transitions[38]. It could be used to investigate the doping dependence of the spectrum in correlated electronic phenomena such as in superconductors, Mott insulators, charge-density-wave materials, and excitonic insulators. It could be used to observe spectral reconstructions in structures with moiré superlattice modulations, including magic-angle twisted bilayer graphene[39,40]. With the addition of circularly polarized light or a spin-resolved spectrometer, it could be applied to electrically controlled magnetic phenomena, such as a half-metal state arising when the Fermi level is tuned between spin-split bands[41].

**Materials and Methods**

**Sample fabrication.** Samples were fabricated by standard micro-mechanical exfoliation and dry transfer[42] with polycarbonate film-based stamping, as detailed in SI section S1.

**Angle resolved photoemission spectra** were acquired at the Spectromicroscopy beamline of the Elettra light source[43]. Linearly polarized light, at 45° to the sample, was focused to a submicrometre diameter spot by a Schwarzschild objective. The energy and momentum of the photoemitted electrons was measured with a hemispherical analyser, at typical instrument resolutions of 50 meV and 0.03 Å$^{-1}$ respectively. The photon energy was 27 eV except for the data in Fig. 1 which was acquired using a photon energy of 74 eV. For imaging, with the light focussed at a fixed spot, the photoelectron intensity on the detector was acquired point by point as the sample was stepped relative to the light spot. In the SPEM images shown, the colour at each pixel corresponds to the integrated photoelectron intensity (over the full detector range of ∼ 3.5 eV and ∼ 15°, corresponding to ∼ 0.6 Å$^{-1}$ at 20 eV) at that point on the sample. Prior to measurement, samples were annealed in ultrahigh vacuum at up to 650 K for a few hours. The sample temperature during measurement was ∼100 K (Figures 2,3 and 4) or ∼105 K (Figure 1 and $MoS_2$, $WS_2$ and $MoSe_2$ data in supplementary information).

**Electronic structure calculations** including spin-orbit interaction were made using the Quantum Espresso DFT package[44] as described previously[10]. Structures were first optimized until forces were smaller than $10^{-4}$ Ry / Bohr for monolayers, and $5\times10^{-4}$ Ry / Bohr for multilayers. Subsequently, the band structures were calculated and sampled on a 12x12 in-plane k-point grid with 800 eV plane-wave energy and 8000 eV charge density cutoffs. To avoid interaction between periodic images, the simulation cell height was 30.0 Å.

**Estimating the energy of the conduction band edge, $E_C$.** The density of states at a single band edge is $g_{2D} = g_s g_v m^*/\hbar^2$, with spin and valley degeneracies $g_s$ and $g_v$, and effective mass $m^*$. For 1L $WSe_2$ the conduction band edges are at the K-points, so $g_v = 2$, and DFT predicts[31] electron mass $m^* \approx 0.3\, m_e$ (the dispersion of the conduction band is not resolvable in the ARPES data here). The band is spin-split by $\approx 40$ meV [31] (also not resolvable) so at moderate carrier concentrations we can take it to be spin-polarised at 100 K, hence $g_s = 1$, and obtain an estimate of $E_F - E_C$ from $n_G = \int_{E_c}^{\infty} F(E) g_{2D}\, dE$, where $F(E)$ is the Fermi-Dirac distribution. For higher carrier concentrations, $n_G > \sim 1.0 \times 10^{13}$ cm$^{-2}$, the band edges at **Q** (for which $g_s = 2$ and $g_v = 6$) as well as the upper spin-split band at **K** will start to populate and $E_F - E_C$ will rise much more slowly with $n_g$. Despite this, for lower carrier concentrations, accurate knowledge of $n_G$ allows determination of $E_C$ to within uncertainty of order 10 meV and gives $E_F - E_C \approx 30$ meV at $n_G = 1.0 \times 10^{13}$ cm$^{-2}$.

**Optical spectroscopy.** Photoluminescence measurements were performed using ∼ 20 μW of linearly polarized 532 nm continuous-wave laser excitation in reflection geometry, with the signal collected by a spectrometer and a Si charge-coupled device, in vacuum in a closed-cycle cryostat.

**Acknowledgements**

The Engineering and Physical Sciences Research Council is acknowledged for support through EP/P01139X/1 and a studentship for NCT (EP/M508184/1). XXia was supported by a University of Warwick studentship. DHC and PVN were supported by US Department of Energy, Office of Basic Energy Sciences, Division of Materials Sciences and Engineering, award DE-SC0002197. PVN and JK were also supported by NSF MRSEC award 1719797. XXu and NPW are supported by Department of Energy, Basic Energy Sciences, Materials Sciences and Engineering Division (DE-SC0018171). NDMH and GCC acknowledge the support of the Winton Programme for the Physics of Sustainability. Computing resources were provided by the Darwin Supercomputer of the University of Cambridge

High Performance Computing Service. GCC acknowledges the support of the Cambridge Trust European Scholarship.

# Visualizing electrostatic gating effects in two-dimensional heterostructures

## Supplementary Information and Materials

### Authors
Paul V Nguyen,[1] Natalie C Teutsch,[2] Nathan Wilson,[1] Joshua Kahn,[1] Xue Xia,[2] Viktor Kandyba,[3] Alexei Barinov,[3] Gabriel C Constantinescu,[4] Nicholas D M Hine,[2] Xiaodong Xu,[1,5*] David H Cobden,[1*] and Neil R Wilson[2*]

### Affiliations
[1]Department of Physics, University of Washington, Seattle, Washington 98195, USA
[2]Department of Physics, University of Warwick, Coventry, CV4 7AL, UK
[3]Elettra - Sincrotrone Trieste, S.C.p.A., Basovizza (TS), 34149, Italy
[4]TCM Group, Cavendish Laboratory, University of Cambridge, 19 JJ Thomson Avenue, Cambridge CB3 0HE, UK
[5]Department of Material Science and Engineering, University of Washington, Seattle, Washington 98195, USA

[*]Email: DHC (cobden@uw.edu ), NRW (Neil.Wilson@warwick.ac.uk), XX (xuxd@uw.edu)

## Table of Contents

**Section S1. Fabrication of a gated heterostructure**

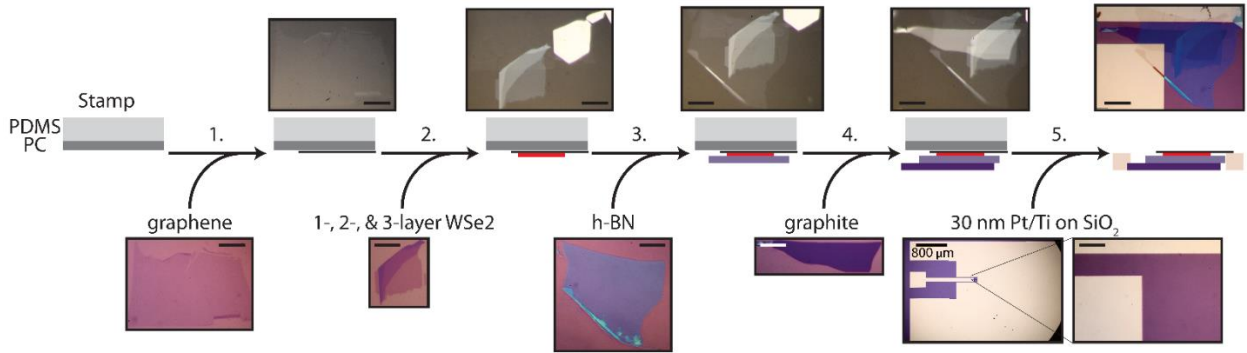

Fig. S1. Fabrication of a graphene, WSe$_2$, hBN, graphite heterostructure. Schematic of fabrication process for the sample in Fig 2. of the main text. Scale bars are 15 µm unless otherwise noted.

Fig. S1 shows schematically the fabrication process, which follows the steps described below.

1.  Graphene was exfoliated at 100 °C onto SiO$_2$ that had been freshly oxygen plasma treated at 50W for 5 minutes[1]. The graphene flake was transferred by dry transfer[2], onto a polycarbonate (PC) on polydimethylsiloxane (PDMS) stamp on a glass slide mounted to a Märzhäuser Wetzlar SM 3.25 motorized micromanipulator controlled by joystick via a MW Tango controller. The manipulator brought stamp and graphene into contact at 130 °C. The substrate was heated to 150 °C, then allowed to cool to 70 °C while the manipulator was set to maintain the relative position between stamp and substrate to prevent PC film delamination. When the substrate reached 70 °C, the manipulator steadily raised the stamp (at $\sim 1\ \mu m/s$) to yield graphene-on-stamp.

2.  Large, thin WSe$_2$ was exfoliated onto SiO$_2$ using the following procedure. A master 'tape' was prepared using scotch tape and a large ($< 5 \times 5$ mm) bulk crystal, which was copied onto the tape with the goal of preserving the large contiguous surfaces of the original crystal and its copies. Immediately after the SiO$_2$ oxygen plasma treatment, a blank piece of tape was placed gently onto the master tape without any intentional pressing or rubbing. With the original tape held down, the second tape was quickly ripped away. The new tape was then exfoliated onto the SiO$_2$ with mild pressure with broad-tipped polyamide/carbon-fibre tweezers. The tape and SiO$_2$ were heated to 120 °C for two minutes and allowed to cool to room temperature. Finally, the tape was peeled away steadily (at $< 1$ mm/s) and at less than 45 ° from the SiO$_2$ surface. A suitable flake was then identified by optical contrast, then aligned to and transferred onto the graphene-on-stamp by the same overall procedure as in step 1, though greater care was taken to bring the graphene-on-stamp and WSe$_2$ into contact steadily using the micromanipulator.

3.  A thin flake of hBN was exfoliated at 100 °C onto SiO$_2$ similarly to graphene in step 1., and its thickness and surface cleanliness was checked by tapping-mode AFM. It was then aligned to and transferred onto the heterostructure-on-stamp as in step 2.

4.  Step 3 was repeated for graphite. Note that the amount of graphite not under hBN was minimized, as the PC film itself does not reliably pick-up graphite.

5. The final SiO$_2$ substrate with 30/5 nm thick Pt/Ti electrodes was prepared by standard e-beam lithography and evaporation, and cleaned by annealing in an Argon (95%)/Hydrogen (5%) atmosphere at 400 °C. The grounding electrode wraps around the gate electrode to minimize stray fields which may deflect photoelectrons. The heterostructure-on-stamp was aligned to and placed onto the substrate at 130 °C. The assembly was heated to up to 175 °C, at which point the PC delaminated from the PDMS when the stamp was raised from the substrate. The PC-on-substrate was torn away by lateral motion, leaving heterostructure-under-PC on the substrate. The PC was removed by a solvent wash in repeated chloroform/IPA baths (with the substrate transferred between baths in a small boat to prevent solvent sloshing which can damage the heterostructure), rinsed in IPA and dried under N$_2$ gas followed by thermal annealing in an Ar/H$_2$ atmosphere at 225 °C for 30 minutes.

**Section S2. Gate dependence of graphene Fermi velocity**

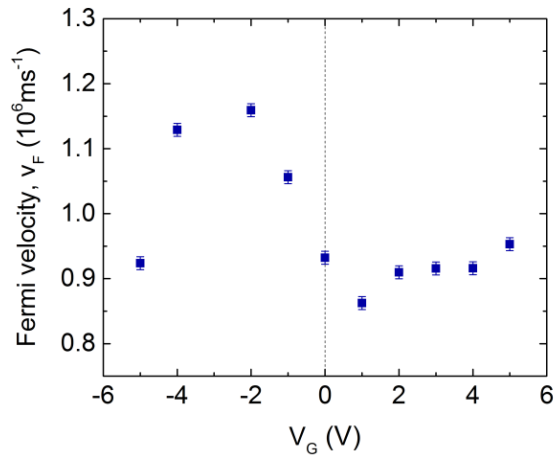

Fig. S2. Dependence of graphene Fermi velocity on gate voltage.

The Dirac point and Fermi velocity were found from $(E, k)$ slices near the graphene K-point, some of which are shown in Fig.1 of the main text, through analysis of the band dispersions. Momentum distribution curves (MDCs, intensity as a function of momentum $I(k)$ at constant energy) were extracted and the position of each side of the Dirac cone found by fitting a Gaussian peak at each band. Repeating this process for each MDC within $E - E_F < 1$ eV the band dispersions as a function of energy and momentum for each side of the Dirac cone were extracted. These are roughly linear; by fitting a straight line to each side, the Dirac point $E_D$ was found from where these lines cross and the Fermi velocity $v_F$ from the slope of the line (where one side was much more intense than the other, the intense side was used to calculate $v_F$ to reduce uncertainty).

The Fermi velocity was roughly constant with gate voltage, Figure S2. It is expected to increase at low carrier concentration ($\sim 1 \times 10^{12}$ cm$^{-2}$),[3,4] but that regime was not probed in this work.

## Section S3. Calculating carrier concentration from $V_G$ and ARPES data

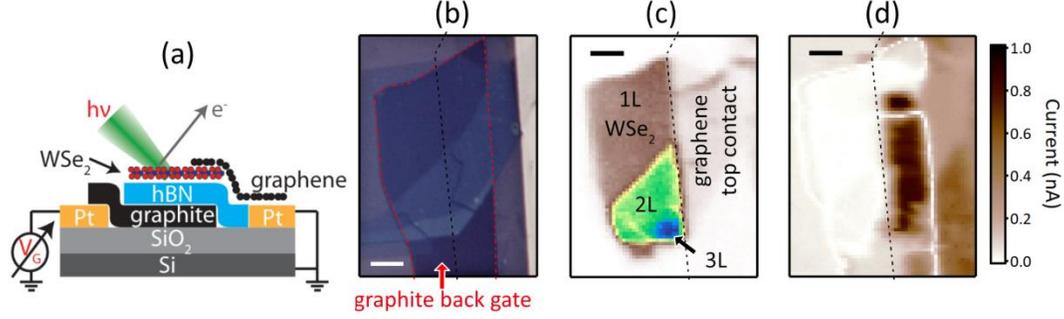

Fig. S3. Photocurrent SPEM map of the WSe₂ heterostructure. (a) Schematic device, with graphene contact grounded and a gate voltage applied to the graphite back gate. (b) Optical and (c) SPEM images of WSe$_2$ device 1 with 1L, 2L and 3L regions identified. (d) Photocurrent image acquired simultaneously to the SPEM image in (c). Scale bars are 5 µm.

The field-effect devices in this work used a thin hBN dielectric to separate the graphite back-gate electrode from an upper 2D material (2DM) layer grounded by a graphene contact, as in Fig. S3a. When the upper 2DM is conducting, this forms a parallel-plate capacitor geometry. The gate-induced change in carrier concentration of the 2DM, $n_G$, can then be calculated from the potential difference across the hBN dielectric using the geometric areal capacitance of the hBN layer $C_g^{hBN} = \frac{\varepsilon_0 \varepsilon_{BN}}{d_{BN}}$, where $\varepsilon_0$ is the relative permittivity of free space, $\varepsilon_{BN} = 4.5 \pm 0.1$ is the out-of-plane (c direction) dielectric constant for hBN and $d_{BN}$ is the thickness of the hBN. The potential drop across the hBN is not defined solely by the applied gate voltage, $V_G$; care must be taken to account for the change in chemical potential in the 2DM, Δµ, as well as any electrostatic potential drop between the 2DM and ground, ΔV. Then $n_G = C_g^{BN}(V_G - \Delta\mu - \Delta V)$. ΔV is dependent on the conductivity of the 2DM, the resistance of the contact between the 2DM and ground electrode, and the photoemission current produced by the µ-ARPES measurement which here was typically ~ 1 nA at $V_G = 0$, as shown in Fig. S3d.

For the gated graphene device shown in Fig.1 of the main text, Δµ and ΔV were measured directly from the photoemission spectra and were both small relative to the applied gate voltage. Δµ is equal to the change in Dirac point energy, $E_D$, which was determined as described in Section S2 above. ΔV was determined from the change in the kinetic energy of the photoemitted electrons at $E_F$ and was at most ~30 meV (the uncertainty on the determination of $E_F$ was typically 20 meV). The dominant contribution to the uncertainty in $n_G$ was the uncertainty in the thickness of the hBN (measured by atomic force microscopy).

For the gated MX$_2$ data, Δµ + ΔV was measured directly from the photoemission spectra from the change in kinetic energy of photoemitted electrons from the upper valence band at **K**, $\Delta E_K = \Delta\mu + \Delta V$. Due to the lower conductivity of the semiconducting MX$_2$, particularly at low gate voltages, and the larger changes in chemical potential, this was a more significant contribution with typically Δµ + ΔV ≈ 1 V. However, once this was corrected for, the dominant contribution o the uncertainty in $n_G$ was again the uncertainty in the thickness $d_{BN}$ of the hBN (~ 5%).

# Section S4. Conduction bands of single layer MoS₂, MoSe₂, and WS₂

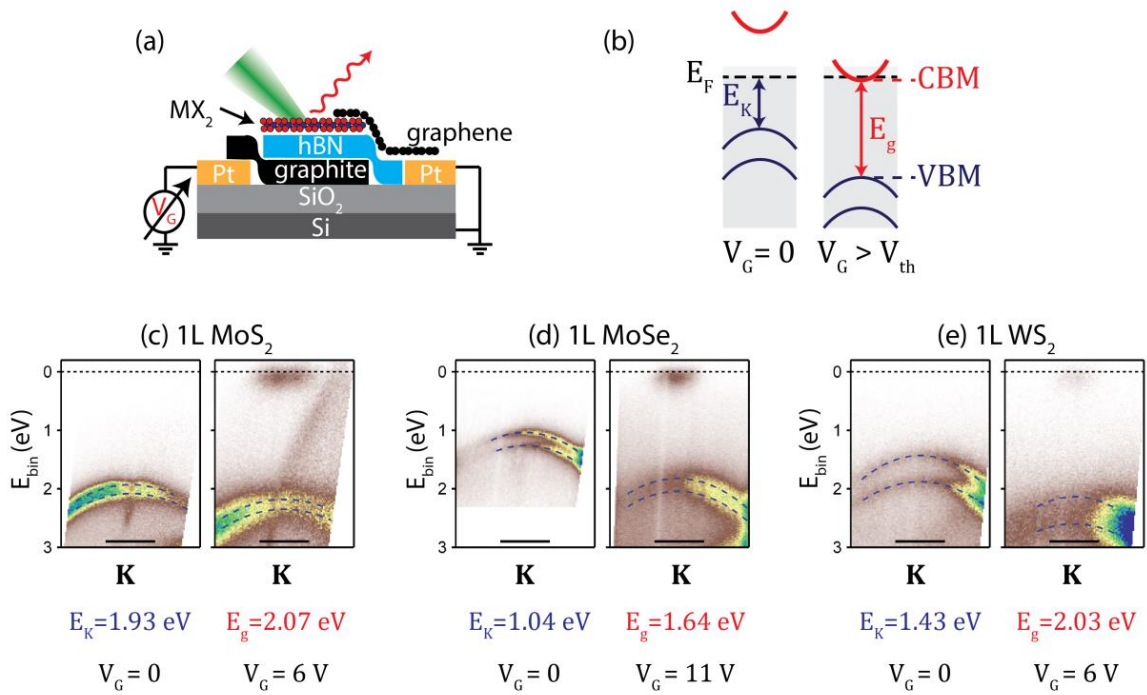

Fig. S4. ARPES of gated monolayer semiconducting MX$_2$. (a) Schematic device, with graphene contact grounded and a gate voltage applied to the graphite back gate. (b) Schematic of the bands at **K** with no gate voltage and with gate voltage exceeding the threshold voltage to bring the conduction band edge below the chemical potential. Energy momentum slices at **K** for 1L 2D semiconductors of MoS$_2$ (c), MoSe$_2$ (d), and WS$_2$ (e). Scale bars are 0.3 Å$^{-1}$.